\documentclass[fleqn,usenatbib]{mnras}

\usepackage[T1]{fontenc}

\DeclareRobustCommand{\VAN}[3]{#2}
\let\VANthebibliography\thebibliography
\def\thebibliography{\DeclareRobustCommand{\VAN}[3]{##3}\VANthebibliography}

\usepackage{graphicx}	
\usepackage{amsmath}	
\usepackage{bm}
\usepackage{caption}
\usepackage{subcaption}

\usepackage{cancel}
\usepackage[dvipsnames]{xcolor}

\usepackage{newtxtext,newtxmath}

\title[Turbulent stress within dead zones]{Turbulent stress within dead zones and magnetic field dragging induced by Rossby vortices}

\author[Raúl O. Chametla et al.]{
Raúl O. Chametla,$^{1}$\thanks{E-mail: raul@sirrah.troja.mff.cuni.cz (ROC)}
Ondrej Chrenko,$^{1}$
Mauricio Reyes-Ruiz$^{2}$
and F. J. Sánchez-Salcedo$^{3}$
\\
$^{1}$Charles University, Faculty of Mathematics and Physics, Astronomical Institute V Hole$\check{s}$ovi$\check{c}$k\'ac 747/2, 180 00 Prague 8, Czech Republic\\
$^{2}$Instituto de Astronomía, Universidad Nacional Autónoma de México, Ensenada, 22800 B.C, Mexico\\
$^{3}$Instituto de Astronomía, Universidad Nacional Autónoma de México, Ciudad Universitaria, Apt. Postal 70-264, C.P. 04510, Mexico City, Mexico
}

\date{Accepted XXX. Received YYY; in original form ZZZ}

\pubyear{2015}

\begin{document}
\label{firstpage}
\pagerange{\pageref{firstpage}--\pageref{lastpage}}
\maketitle

\begin{abstract}
By means of three dimensional resistive-magnetohydrodynamical models, we study the evolution of the so-called dead zones focused on the magnitude of the Reynolds and Maxwell stresses. We consider two different types of static resistivity radial profiles which give rise to an intermediate dead zone or an intermediate active zone. As we are interested in analyzing the strength of angular momentum transport in these intermediate regions of the disc, we use as free parameters the radial extent of the intermediate dead ($\Delta r_\mathrm{idz}$) or active ($\Delta r_\mathrm{iact}$) zones, and the widths of the inner ($H_{b_1}$) and outer ($H_{b_2}$) transitions. We find that regardless of the width or radial extent of the intermediate zones, Rossby wave instability (RWI) develops 
at these transition boundaries, leading to the emergence of vortices and spiral waves. In the case of an intermediate dead zone, when $H_{b_1}\,,H_{b_2}\leq0.8$, the vortices are almost completely confined to the dead zone. Remarkably, we find that the formation of vortices at the inner transition can drag magnetic field lines into the dead zone stirring up the region that the vortex covers (reaching an $\alpha\approx10^{-2}$ value similar to that of an active zone). Vortices formed in the outer transition only modify the Reynolds stress tensor. Our results can be important to understanding angular momentum transport in poorly ionized regions within the disc due to magnetized vortices within dead zones. 
\end{abstract}




\section{Introduction}
\label{sec:intro}

A few decades ago when the first exoplanet was discovered \citep[][]{MQ1995}, the interest in planetary migration increased notably since it could provide a point of comparison with theoretical models which attempts to explain the formation and migration of planets around Sun-like stars \citep[see for instance:][]{Kley_etal2012,Baruteau_etal2014,Paardekooper_etal2022}. However, to date no completely conclusive results have been obtained. One important issue lies in understanding how angular momentum transport occurs in protoplanetary discs. It is widely accepted that the turbulence generated in the disc by the magnetorotational instability \citep[MRI; see][for details of this instability]{BH1991} is the major source of turbulent viscosity, which results in outward momentum transport and mass accretion through the disc. 

However, for MRI to operate efficiently, the gas must be sufficiently ionized so that it can fully couple to the magnetic field. In the region very close to the star, due to the high temperatures (above 1000 $\mathrm{K}$), collisional ionization produces such coupling \citep[][]{U1983,UN1988}. Outside this inner region, the ionization process is driven by non-thermal ionization processes, such as ionization by X-rays and cosmic rays from young stars and interstellar space, respectively \citep[][]{Glas2004,U1983,UN1988}, as well as ionization by the decay of radionuclides within the gas \citep{UN1981}. Other sources of ionization have recently been explored such as a nearby supernova explosion, the corona of the protoplanetary disc, and the ionization from a very young star \citep[see][]{TD2009}.

Despite the different ionization processes existing in protoplanetary discs, it has been argued that there is a region close to the mid-plane of the disc, so-called the dead zone, where the MRI is suppressed \citep[][]{Gammie1996,Sano2000,Frog2002,IN2006,TSD2007,TD2009}. The shape and size of the dead zone is defined mainly by Ohmic, Hall and ambipolar diffusion \citep[see][and references therein]{Natalia2013}. 

On the other hand, within the dead zone there may be disturbances generated from the active vertical and radial zones of the disc. Shearing-box magnetohydrodynamical (MHD) stratified disc models including Ohmic resistivity as well as time-dependent ionization chemistry \citep{TSD2007}, show that 
turbulent mixing of free charges can lead to a coupling
between the magnetic field and the otherwise dead zone.
\citet{OMM2007} find that turbulence in the upper and lower active layers can excite density fluctuations in the dead zone. Three-dimensional simulations of magnetized non-stratified discs show that in the radial positions where transitions between the active zone and the dead zone occur, spiral density waves can be excited \citep{Lyra_Mac2012,Lyra_etal2015,ChCh2023}. These spiral waves are the result of the formation of a vortex at the pressure maximum in the transition region due to the Rossby wave instability (RWI) \citep[see][]{Lovelace_etal1999,Li_etal2000,Li_etal2001} and can produce an important change in the rate of accretion including a similar level of Reynolds stress to that of the active zone \citep{Lyra_Mac2012,Ryan2016}. In addition, the formation of vortices at the edges of the transition zones can also produce different morphologies in the protoplanetary discs, depending on the radial width of the dead zone \citep{Ryan2016}. Here we are interested in studying the effect on the Maxwell and Reynolds stresses within the dead zone, due to the formation of vortices at the dead zone edges as a function of the width of the transitions, and the radial extension of the dead zone itself. 

The paper is laid out as follows. In Section \ref{sec:physical}, we present the physical model, code and numerical setup used in our 3D-MHD simulations. In Section \ref{sec:results}, we present the results of our numerical models. We present a brief discussion in Section \ref{sec:discussion}. Concluding remarks can be found in Section \ref{sec:conclusions}.

\begin{figure}
\includegraphics[width=0.45\textwidth]{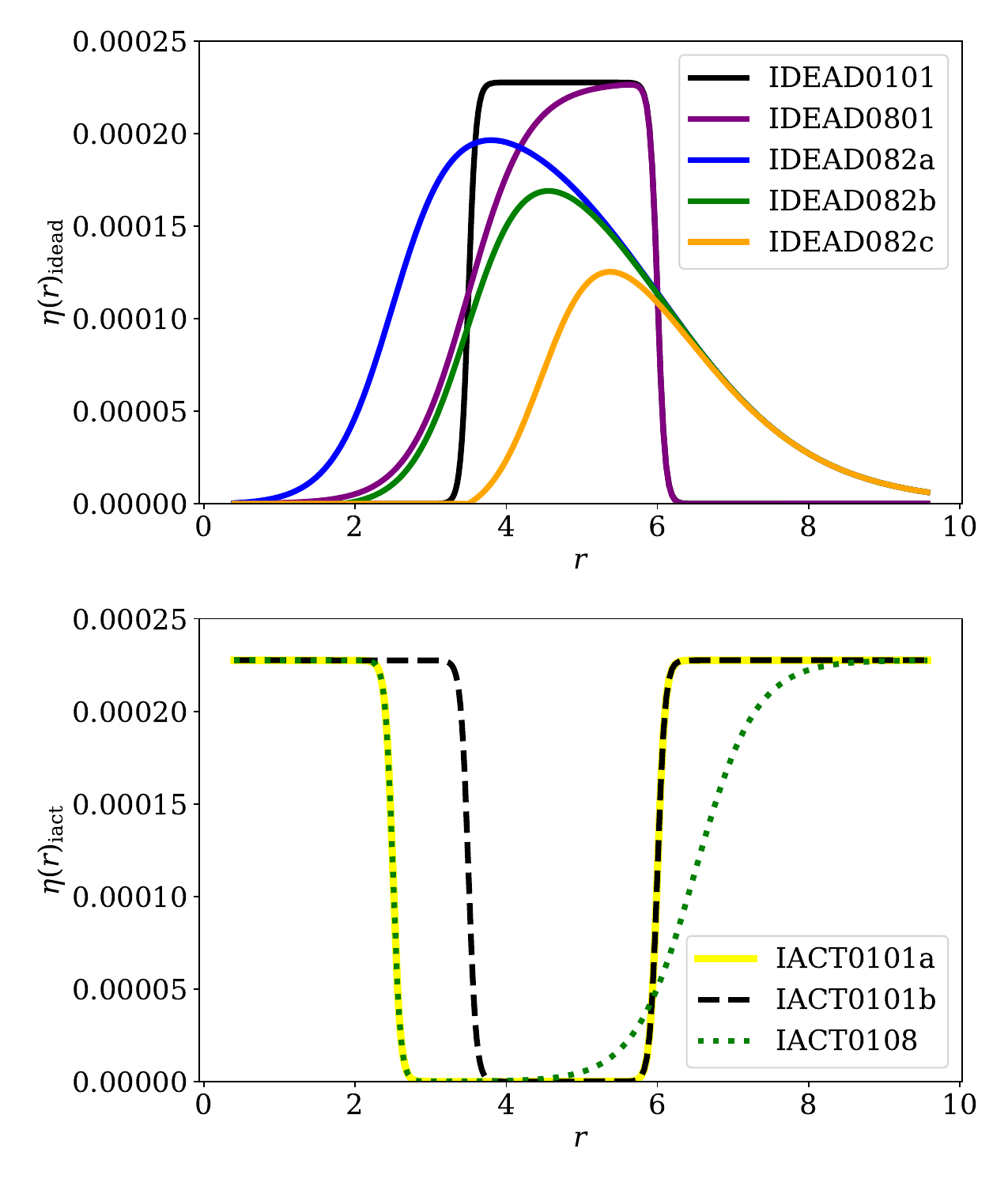}
 \caption{\textit{Top.} Radial resistivity profiles given by Eq. (\ref{eq:eta_idead}) which generate an intermediate dead zone. \textit{Bottom.} Radial resistivity profiles given by Eq. (\ref{eq:eta_iact}) that describe an intermediate active zone. In both cases, the profiles depend on the widths $H_{b_1}$ and $H_{b_2}$ of the internal and external transitions, respectively, given in Table \ref{tab:initial_conditions}.}
\label{fig:rprofiles}
\end{figure}

\begin{table}
\caption{Parameters of the numerical models.}
    \label{tab:initial_conditions}
\begin{tabular}{ p{1.3cm} p{0.7cm} p{0.7cm} p{0.7cm} p{0.7cm} p{0.7cm} p{0.7cm}}
 \hline
 Model & $\tilde{r}_1$ & $\tilde{r}_2$ & $H_{b_1}$ & $H_{b_2}$ & $r_{1}$ & $r_{2}$\\
 \hline
 IACT0101a & 2.5 & 6 & 0.1 & 0.1 & 2.55 & 5.88 \\
 IACT0101b & 3.5 & 6 & 0.1 & 0.1 & 3.58 & 5.88\\
 IDEAD0101 & 3.5 & 6 & 0.1 & 0.1 & 3.42 & 6.12\\
 IACT0108 & 2.5 & 6.5 & 0.1 & 0.8 & 2.55 & 5.55\\
 IDEAD0801 & 3.5 & 6 & 0.8 & 0.1 & 3.0  & 6.12 \\
 IDEAD082a & 2.5 & 6 & 0.8 & 2  & 2.26 & 9.15\\
 IDEAD082b & 3.5 & 6 & 0.8 & 2  & 3.08 & 9.15\\
 IDEAD082c & 4.5 & 6 & 0.8 & 2  & 4.1 & 9.15\\
 
 \hline
\end{tabular}
\end{table}

\begin{figure*}
    \includegraphics[scale=0.5]{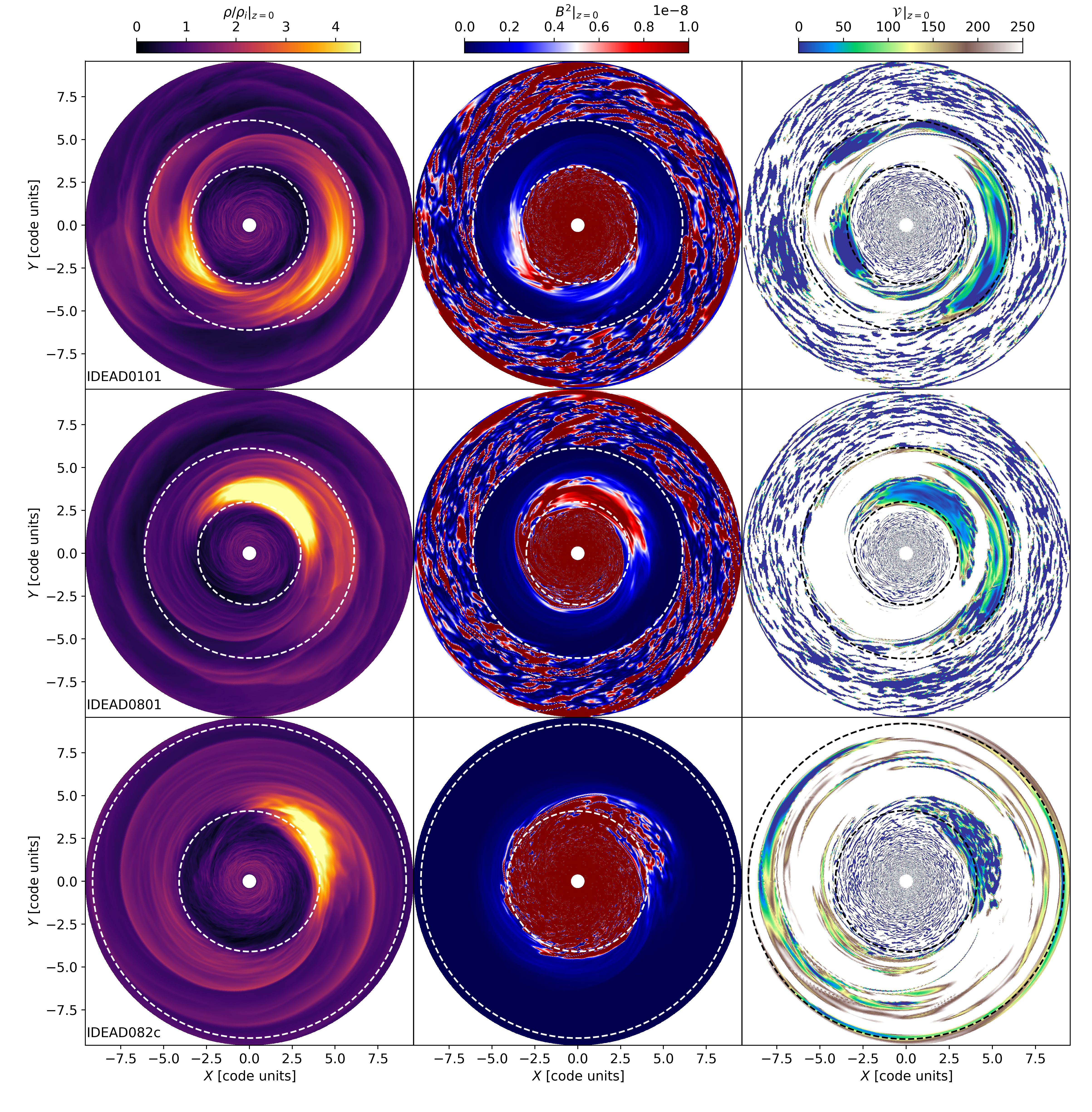}
  \caption{Maps of the normalized gas density $\rho/\rho_i$, squared modulus of the magnetic field $B^2$, and the gas vortensity $\mathcal{V}$ at $z=0$, for the IDEAD0101 (first row), IDEAD0801 (second row) and IDEAD082c (third row) models. The dashed circles delineate the (intermediate) dead zone.}
  \label{fig:DBV}
\end{figure*}

\section{Physical Model} 
 \label{sec:physical}
We have carried out 3D global unstratified MHD simulations of a gaseous disc with a stationary Ohmic resistivity profile. The physical model and numerical setup used in this study basically follow those in \citet{ChCh2023}, which we briefly recall for convenience.

\begin{figure*}
  \begin{subfigure}[b]{0.24\textwidth}
    \includegraphics[scale=0.55]{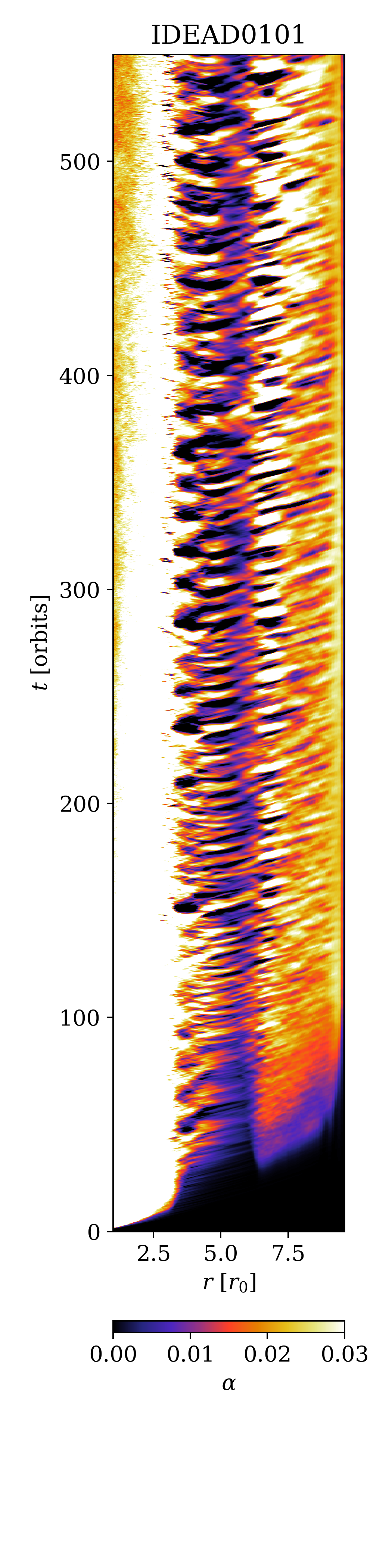}
  \end{subfigure}
  \hfill
  \begin{subfigure}[b]{0.24\textwidth}
    \includegraphics[scale=0.55]{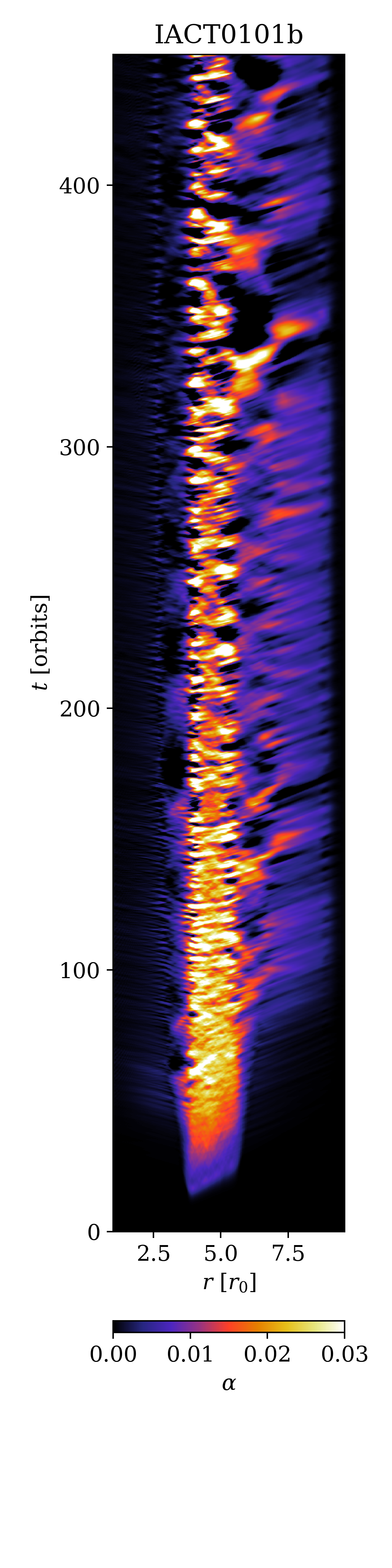}
  \end{subfigure}
  \hfill
  \begin{subfigure}[b]{0.24\textwidth}
    \includegraphics[scale=0.55]{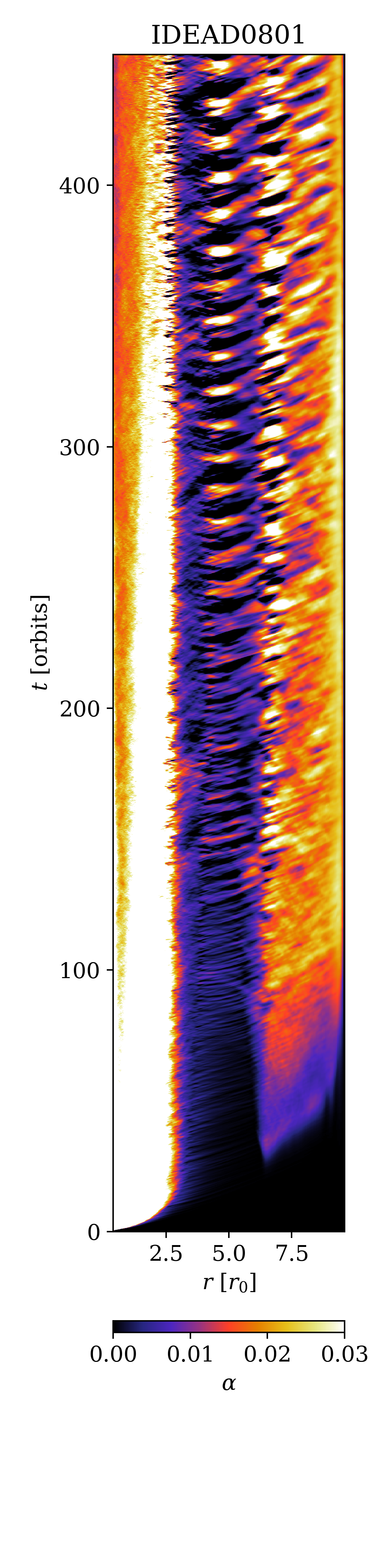}
  \end{subfigure}
  \hfill
  \begin{subfigure}[b]{0.24\textwidth}
    \includegraphics[scale=0.55]{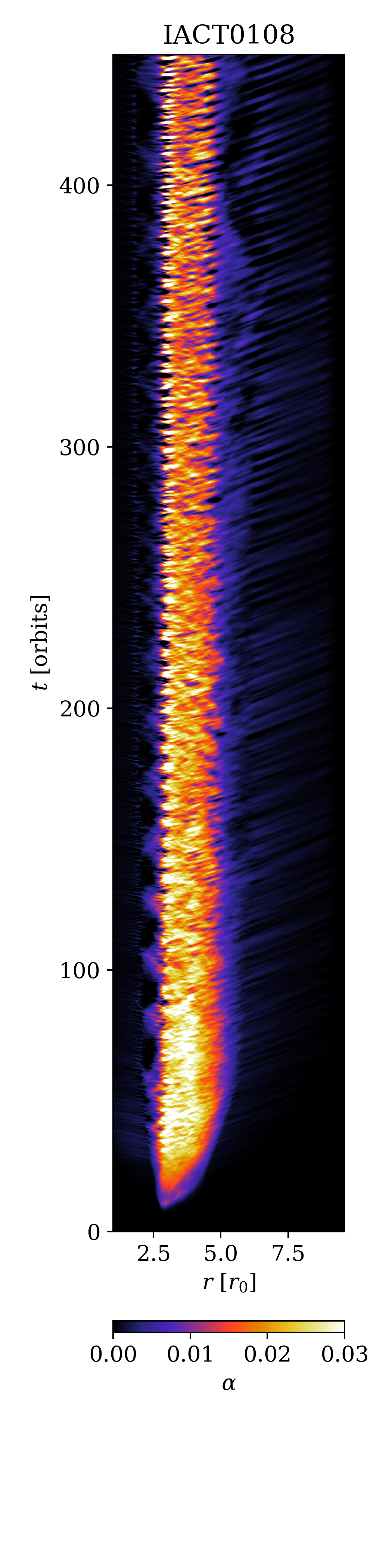}
  \end{subfigure}
  \caption{Temporal evolution of $\alpha$, given by Eq. (\ref{eq:alpha}), for four different models (see Table \ref{tab:initial_conditions}). It can be seen that depending on the widths of the transitions ($H_{b_1}$ or $H_{b_2}$), the magnitude of $\alpha$ changes in the dead zones, within the first 100 orbits, in both the IDEAD and IACT models. Furthermore, this change in $\alpha$ is the result of the development of RWI at the edges of the dead zones (see text for details).}
  \label{fig: aRt}
\end{figure*}

We use a reference frame centred on the star and rotating with the angular frequency $\mathbf{\Omega}=\Omega\mathbf{\hat{k}}$, where $\Omega\equiv\sqrt{GM_{\star}/r_0^3}$ is the Keplerian angular velocity at $r=r_0$ (with $r_0$ is radius of reference), $M_{\star}$ is the stellar mass and $\mathbf{\hat{k}}$ is a unit vector along the rotation axis. The unit of time adopted when discussing the results is $T_0=2\pi/\Omega$. 
Magnetohydrodynamical equations describing the gas flow in 3D-MHD discs are given by the equation of continuity
\begin{equation}
\frac{\partial\rho}{\partial t}=-(\mathbf{v}\cdot\mathbf{\nabla})\rho-\rho\mathbf{\nabla}\cdot\mathbf{v} \,,
\label{eq:continuity}
\end{equation}
the gas momentum equation
\begin{equation}
\begin{split}
\frac{\partial\mathbf{v}}{\partial t}=-(\mathbf{v}\cdot\mathbf{\nabla})\mathbf{v}-\frac{1}{\rho}\mathbf{\nabla} p-\mathbf{\nabla}\Phi+\frac{\mathbf{J}\times\mathbf{B}}{\rho}
&\\ -\mathbf{\Omega}\times(\mathbf{\Omega}\times\mathbf{r})-2\mathbf{\Omega}\times\mathbf{v}\,,
\label{eq:momentum}
\end{split}
\end{equation}
and the induction equation
\begin{equation}
\frac{\partial \mathbf{B}}{\partial t}=\nabla\times\left[\mathbf{v}\times\mathbf{B}-\eta\mathbf{J} \right],
\label{eq:induction}
\end{equation}
where $\rho$ is the gas density, $\mathbf{v}$ is the gas velocity, $\Phi$ is the gravitational potential, and $p$ is the pressure given as
\begin{equation}
p=\rho c_s^2 \,,
\label{eq:pressure}
\end{equation}
with $c_s$ the sound speed. 
We take $c_s=0.1r\Omega_\mathrm{Kep}$ everywhere in the disc, so that if the disc were stratified (which is not)
it would have a scale height $H\equiv 0.1r$.
At $t=0$, we add uniformly distributed noise to the velocity components, cell by cell, equal to $10^{-2}c_s$.

The gravitational potential $\Phi$ is given by
\begin{equation}
\Phi=\Phi_S+\Phi_{\mathrm{ind}},
 \label{eq:potential}
\end{equation}
where
\begin{equation}
\Phi_S=-\frac{GM_\star}{r},
 \label{eq:Star_potential}
\end{equation}
is the stellar potential with $G$ the gravitational constant, and 
\begin{equation}
\Phi_{\mathrm{ind}}= G\int_V\frac{dm(\mathbf{r'})}{r'^3}\textbf{r}\cdot\textbf{r}' \, ,
\label{eq:ind_pot}
\end{equation}
is the indirect potential arising from the gravitational force of the disc, respectively.

In the induction equation, $\mathbf{J}=\mu_0^{-1}\mathbf{\nabla}\times\mathbf{B}$ is the current density and $\eta$ is the Ohmic resistivity. 
We will consider two cases; an intermediate dead ring between two
active regions (models denoted by the IDEAD tag), and an intermediate active 
ring between two dead regions (models tagged as IACT). In the first scenario, 
the Ohmic resistivity is given by 
\begin{equation}
\eta(r)_\mathrm{idead}=\frac{\eta_0}{2}\left[\tanh\left(\frac{r-\tilde{r}_1}{H_{b_1}}\right)-\tanh\left(\frac{r-\tilde{r}_2}{H_{b_2}}\right)\right],
\label{eq:eta_idead}
\end{equation}
where $\tilde{r}_{1}$ and $\tilde{r}_{2}$ are the radial edges of the dead ring, provided that
$H_{b_1}$ and $H_{b_2}$ are sufficiently small
(see also \citet{Lyra_etal2015} and \citet{ChCh2023}). In the
second scenario
\begin{equation}
\eta(r)_\mathrm{iact}=\eta_0-\frac{\eta_0}{2}\left[\tanh\left(\frac{r-\tilde{r}_1}{H_{b_1}}\right)-\tanh\left(\frac{r-\tilde{r}_2}{H_{b_2}}\right)\right].
\label{eq:eta_iact}
\end{equation}
Again, if $H_{b_1}$ and $H_{b_2}$ are much less than $r_{0}$, then $\tilde{r}_{1}$ and 
$\tilde{r}_{2}$ are the radial locations of the edges of the active ring. For very smooth transitions, i.e. if $H_{b_1}$ and/or $H_{b_2}$ are comparable to $r_{0}$, the actual location of the dead/active zone boundaries is determined by the radii $r_{1}$ and $r_{2}$ where the magnetic Reynolds 
number (or Elsasser number), $\mathcal{R}_{m}\equiv v_{A}^{2}/(\eta \Omega_{\rm Kep})$,
is smaller than $1$.

In all cases we take $\eta_0=2.27\times10^{-4}$ as the maximum resistivity value maintained constant over time in the dead zones. Table \ref{tab:initial_conditions} provides the values of $\tilde{r}_{1}$, $\tilde{r}_{2}$,
$H_{b_1}$, $H_{b_2}$ and the corresponding $r_{1}$ and $r_{2}$ for the models explored here, whereas Figure \ref{fig:rprofiles} shows the radial
profiles of $\eta$ as given by Eqs. (\ref{eq:eta_idead}) and (\ref{eq:eta_iact}).
Note that for some values of $H_{b_1}$ or $H_{b_2}$ (when they are sufficiently large), $\eta_{\mathrm {idead}}$ may be negative at some locations.  Since this is unphysical, we take $\eta_\mathrm{idead}=0$ at the intervals where it would be otherwise negative.

The prescription of diffusivity given by equations (\ref{eq:eta_idead}) and (\ref{eq:eta_iact}) is adopted following previous similar studies \citep[e.g.,][]{Lyra_etal2015} in order to isolate the novel results we analyze. It is also justified on account of computational efficiency but it obviously precludes the study of the effects of the evolving disc properties on the location and ionization degree in the dead zones. Future studies should include the feedback of changes in disc density \citep[as discussed for example in][]{FS2003} and temperature  \citep[see][]{Faure2014}, on magnetic diffusivity.

To numerically solve Eqs (\ref{eq:continuity})-(\ref{eq:pressure}), we use the publicly available 
code {FARGO3D\footnote{https://fargo3d.bitbucket.io/intro.html}}
\citep[][]{Benitez_Masset2016} with MHD-orbital advection enabled \citep[see][for details]{Masset2000,Benitez_Masset2016}
in cylindrical coordinates. The FARGO3D code solves the hydrodynamic equations with a time-explicit method, using operator splitting and upwind techniques on an Eulerian mesh. The update of the magnetic field governed by the induction equation (Eq. \ref{eq:induction}) is done by the method of characteristics \citep[][]{Stone_Norman1992} and the constrained transport method \citep[][]{Evans_Hawley1988} is used to preserve the divergence-free property of the magnetic field.

\subsection{Set-up and boundary conditions} \label{sec:model}
The initial gas density follows a power law 
\begin{equation}
\rho_{i}(r)=\rho_0\left(\frac{r}{r_0}\right)^{-q_\rho}.
\label{eq:rho}
\end{equation}
Our MHD simulations have a a net vertical magnetic field. We impose that two MRI wavelengths are contained in the 
vertical extension of the disc, i.e. $2\lambda_{\rm MRI}= L_{z}$, where $\lambda_{\rm MRI}= 2\pi v_{A}/\Omega_{\mathrm{Kep}}(r)$, 
$L_{z}$ the vertical extent of the computational domain and $v_A\equiv B/\sqrt{\mu_0\rho}$ is the Alfvén speed \citep[see][]{Lyra_Mac2012,Lyra_etal2015}. The resulting unperturbed radial profile of the vertical
component of the magnetic field is 
\begin{equation}
B=B_0\left(\frac{r}{r_0}\right)^{-(\frac{3+q_\rho}{2})}
    \label{eq:Bfield}
\end{equation}
(see Appendix \ref{ap:appendixA} for details). In all our models, we fix the value of $q_\rho$ at $1.5$, and the plasma parameter $\beta=2c_s^2/v_A^2$ is set to $10^3$ at $r=r_0$.

\begin{figure}
\includegraphics[width=0.45\textwidth]{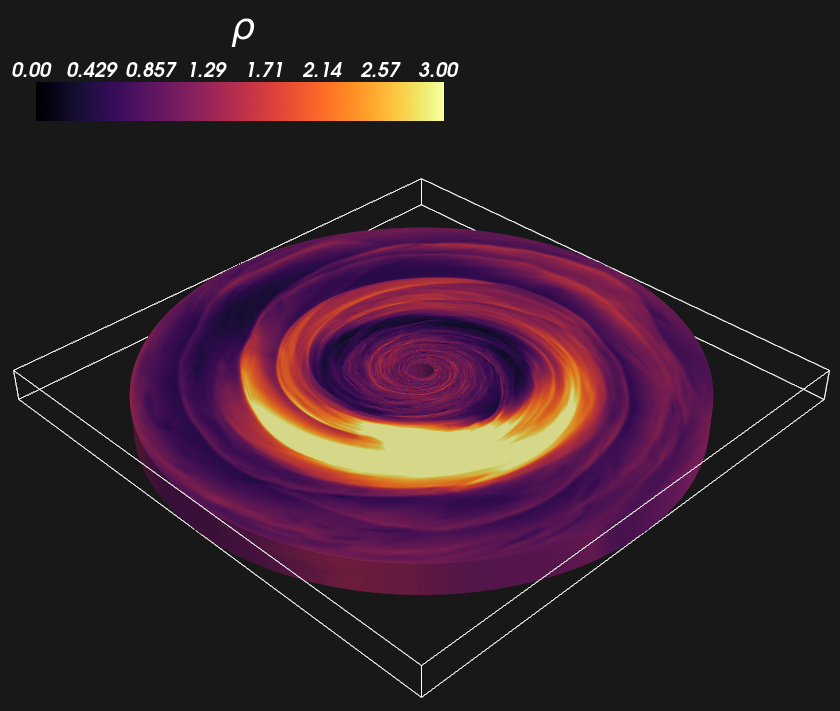}
 \caption{Gas density at $t=550$ orbits in model IDEAD0101.}
\label{fig:dens}
\end{figure}
\begin{figure}
\includegraphics[width=0.45\textwidth]{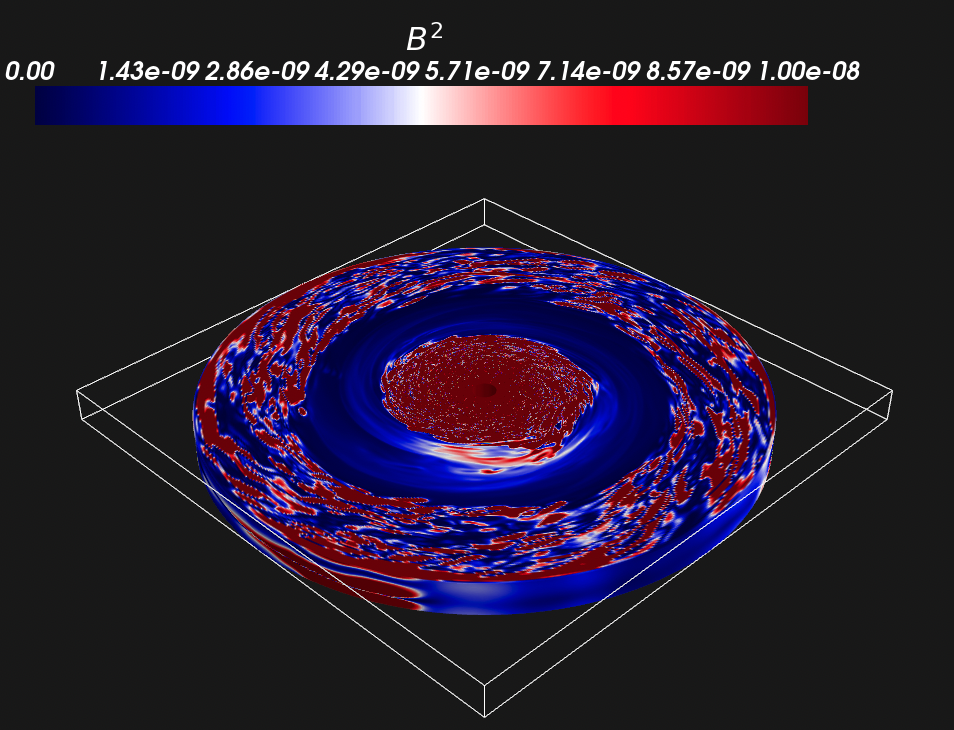}
 \caption{Magnetic field strength at $t=550$ orbits in model IDEAD0101.}
\label{fig:magnetic}
\end{figure}

\begin{figure*}
  \begin{subfigure}[b]{0.49\textwidth}
    \includegraphics[scale=0.5]{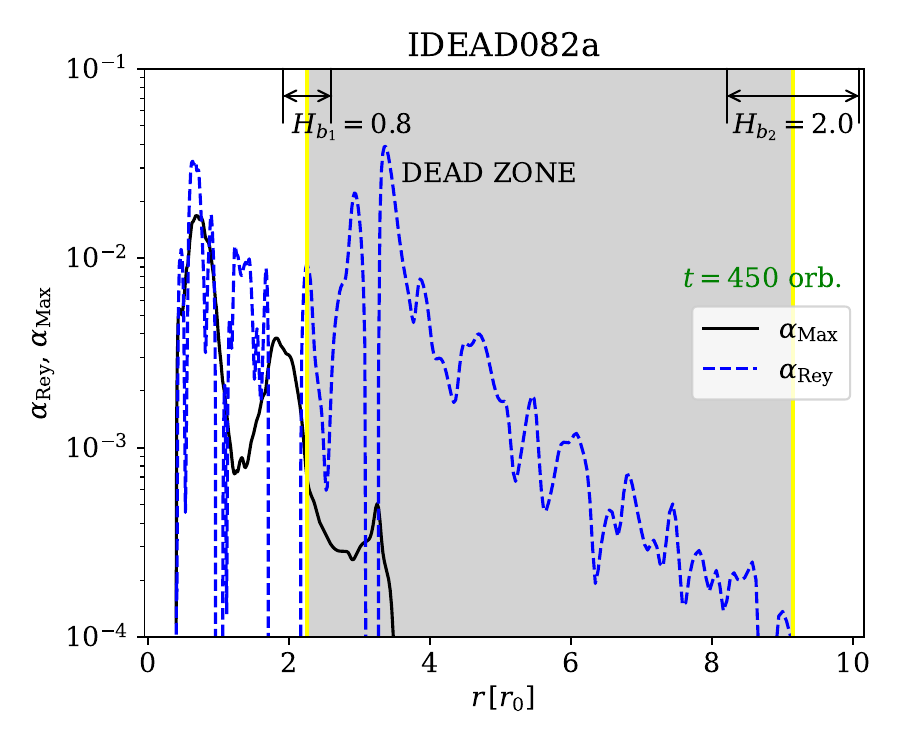}
  \end{subfigure}
  \hfill
  \begin{subfigure}[b]{0.49\textwidth}
    \includegraphics[scale=0.5]{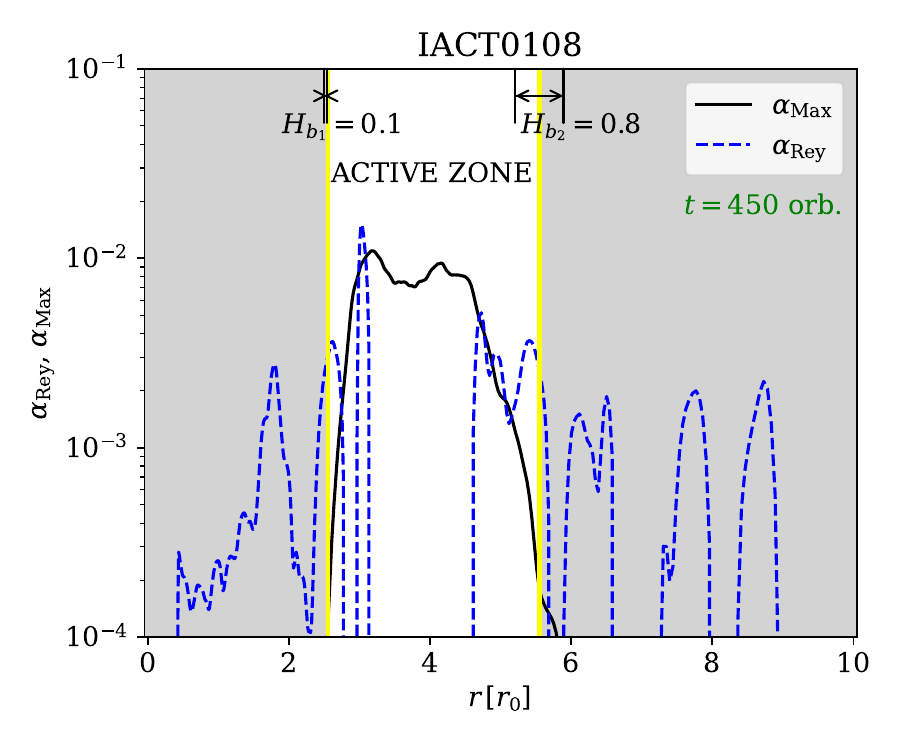}
  \end{subfigure}

   \hfill
  \begin{subfigure}[b]{0.49\textwidth}
    \includegraphics[scale=0.5]{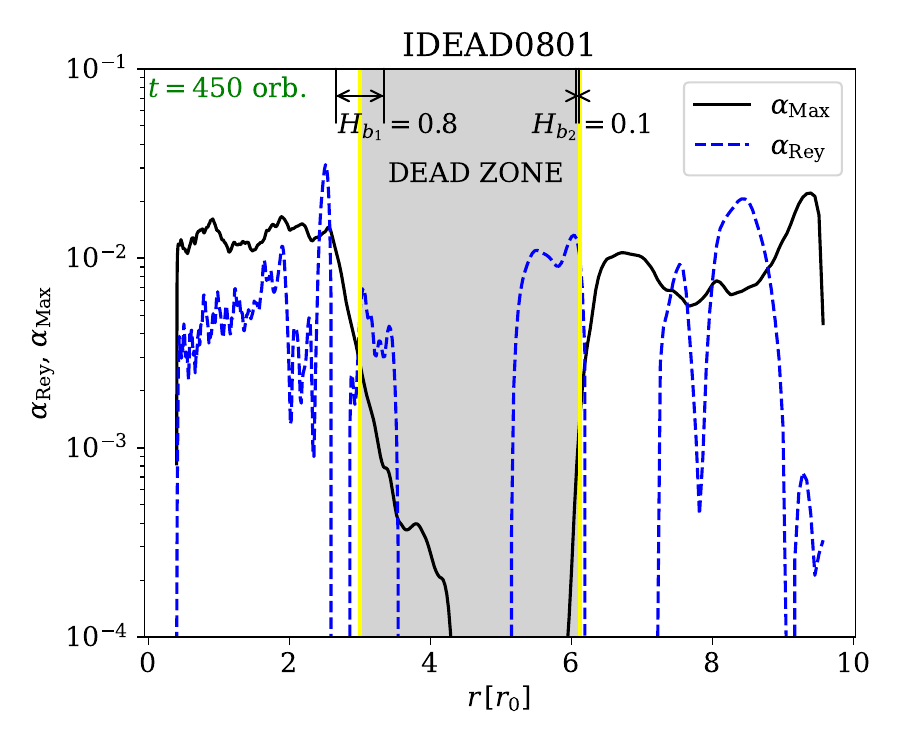}
  \end{subfigure}
  \hfill
  \begin{subfigure}[b]{0.49\textwidth}
    \includegraphics[scale=0.5]{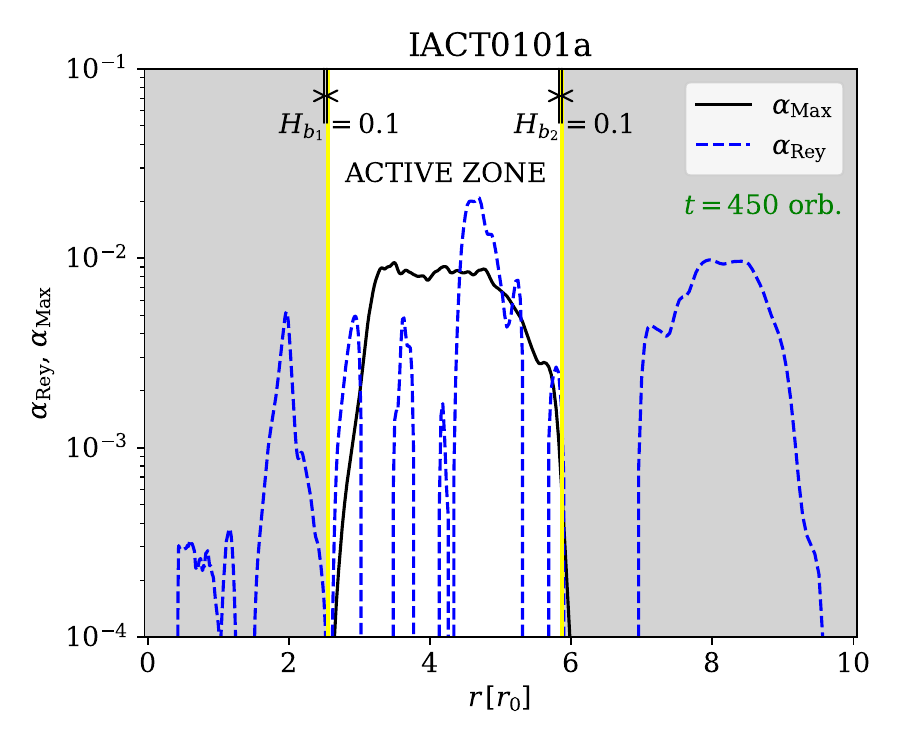}
  \end{subfigure}
  \hfill
  \begin{subfigure}[b]{0.49\textwidth}
    \includegraphics[scale=0.5]{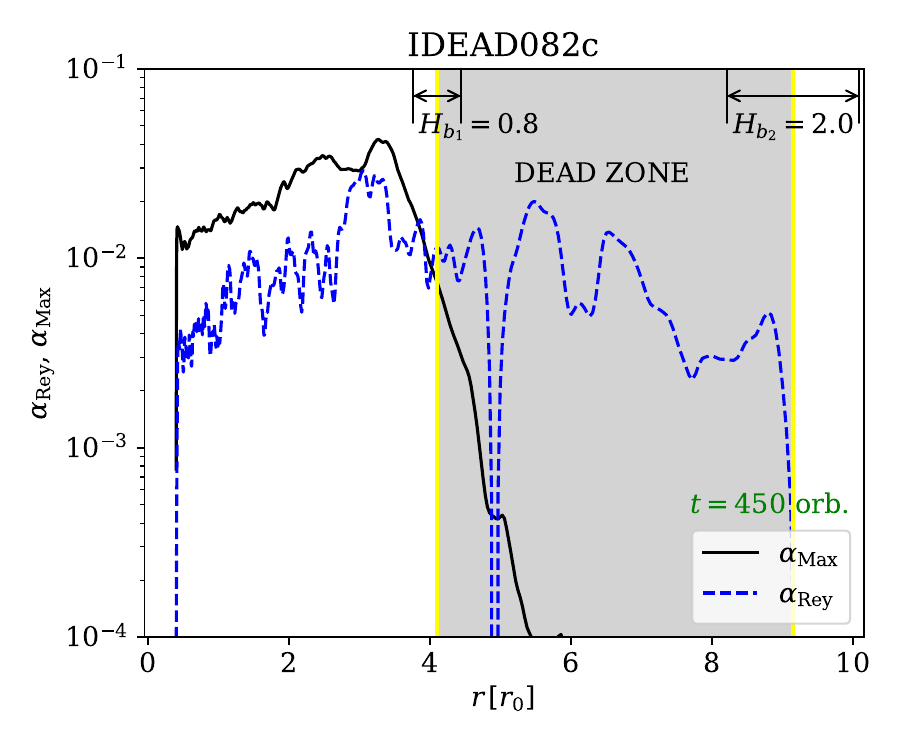}
  \end{subfigure}
  \hfill
  \begin{subfigure}[b]{0.49\textwidth}
    \includegraphics[scale=0.5]{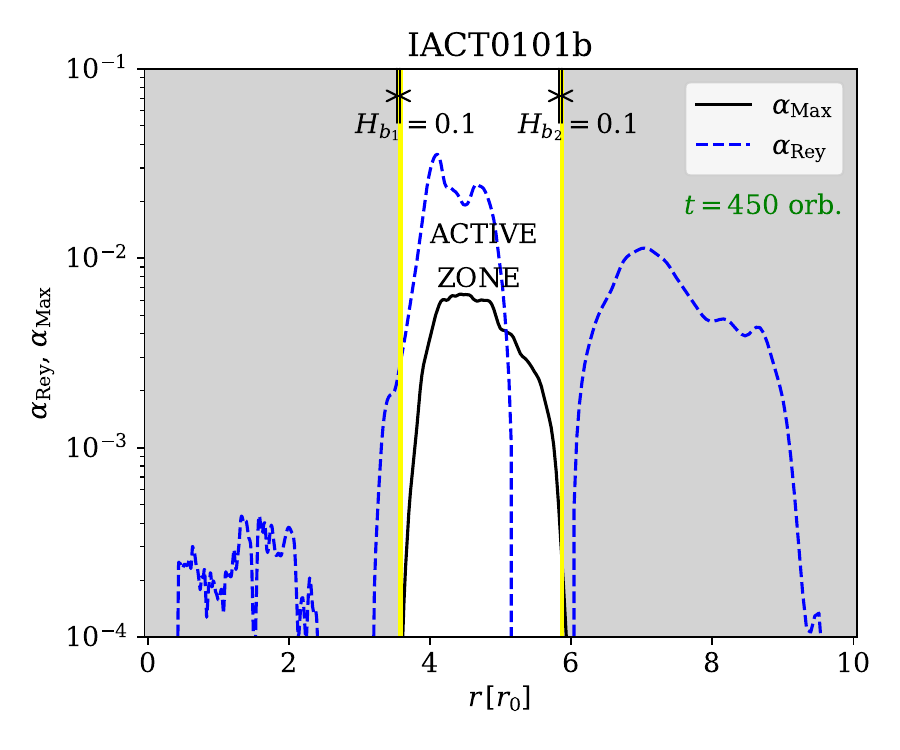}  
  \end{subfigure}
  
  \caption{Radial profiles of $\alpha_\mathrm{Rey}$ and $\alpha_\mathrm{Max}$ for six different models with resistivity transitions given by Eqs (\ref{eq:eta_idead}) and (\ref{eq:eta_iact}). Shaded regions correspond to dead zones, whereas unshaded regions indicate active zones. The yellow vertical lines mark the radial positions of $r_1$ and $r_2$. The line segments at the top of
  each panel represent the widths $H_{b_1}$ and $H_{b_2}$ of the internal and external transitions centred at $r_1$ and $r_2$, respectively.}
  \label{fig: aRM}
\end{figure*}

For extent of our computational domain is $r=[0.4,9.6]r_0$, $\phi=[-\pi,\pi]$ and $z=[-0.1,0.1]r_{0}$ with a radial logarithmic spacing and uniform in both azimuthal and vertical directions\footnote[1]{The vertical extent of the disc model used in this study has as its main objective to guarantee that the MRI is correctly resolved, and to be able to run the simulations at a longer orbital time at a reasonable computational cost.}. The number of zones in each direction is $(N_r,N_\phi,N_z)=(518,1024,64)$.

We use periodic boundary conditions in the vertical direction. To avoid reflections at the radial boundaries of our computational domain,
we use damping boundary conditions as in \citet{deVal2006} for the gas density and in the velocity components. The width
of the inner damping ring being $3.9\times 10^{-2}r_0$ and that of the
outer ring being $8.54\times 10^{-1}r_0$. The damping timescale at
the edge of each damping ring equals $1/20^{\mathrm{th}}$ of the local orbital period. Since the formation of vortices occurs in the intermediate region of the disc (either in the dead zone or in the intermediate active zone) the choice of the width of the damping rings and the damping timescale does not drastically modify our results.

\subsection{Diagnostics}
\label{sec:diag}
To describe the magnitude of the turbulence generated in each of our models we calculate the mass-averaged value of the quantity $Q$
over the azimuthal and vertical directions, by the expression
\begin{equation}
\overline{Q}(r,t)=\dfrac{\int \rho Q\,dz\,d\phi}{\int \rho \,dz \,d\phi}.
    \label{eq:quantities}
\end{equation}
For instance, we calculate the mass-averaged values of Maxwell tensor $\alpha_\mathrm{Max}$ as
\begin{equation}
\alpha_\mathrm{Rey}(r,t)=\dfrac{\int \rho \,\delta v_r\,\delta v_\phi\,dz\,d\phi}{c_{s}^{2}(r)\int \rho \,dz \,d\phi},
    \label{eq:alphaR}
\end{equation}
where $\delta v_r= v_r-\bar{v}_r$ and $\delta v_\phi= v_\phi-\bar{v}_\phi$. On the other hand, $\alpha_\mathrm{Max}$ will be
computed as
\begin{equation}
\alpha_\mathrm{Max} (r,t)=\dfrac{\int \rho \,(\frac{-B_r B_\phi}{4\pi\rho})\,dz\,d\phi}{c_{s}^{2}(r)\int \rho \,dz \,d\phi}.
  \label{eq:alphaM}
\end{equation}
Then the Shakura and Sunyaev parameter is given by
\begin{equation}
\alpha=\alpha_\mathrm{Rey}+\alpha_\mathrm{Max}.
    \label{eq:alpha}
\end{equation}

The vortex formation is analyzed by the vortensity $\mathcal{V}\equiv\zeta_z/\Sigma$, where $\zeta_z$ is the vertical component of the vorticity vector defined, in a rotating frame, as
\begin{equation}
\bm{\zeta}\equiv\nabla\times\mathbf{u}+2\mathbf{\Omega}.
 \label{eq:vorticity_}
\end{equation}

\section{Results}
\label{sec:results}

As said in Section \ref{sec:physical}, we have performed simulations of an intermediate dead zone (radially surrounded by two active zones; models IDEAD), and simulations of an intermediate active zone (radially surrounded by two dead zones; models IACT). We have explored different combinations
of $r_{1}$, $r_{2}$, $H_{b_1}$ and $H_{b_2}$ (see Table {\ref{tab:initial_conditions}). We will focus first on the distribution and structures in the gas density and magnetic field in the IDEAD models. Later, we will discuss models with an intermediate active zone (IACT models).

\begin{figure*}
    \includegraphics[scale=0.5]{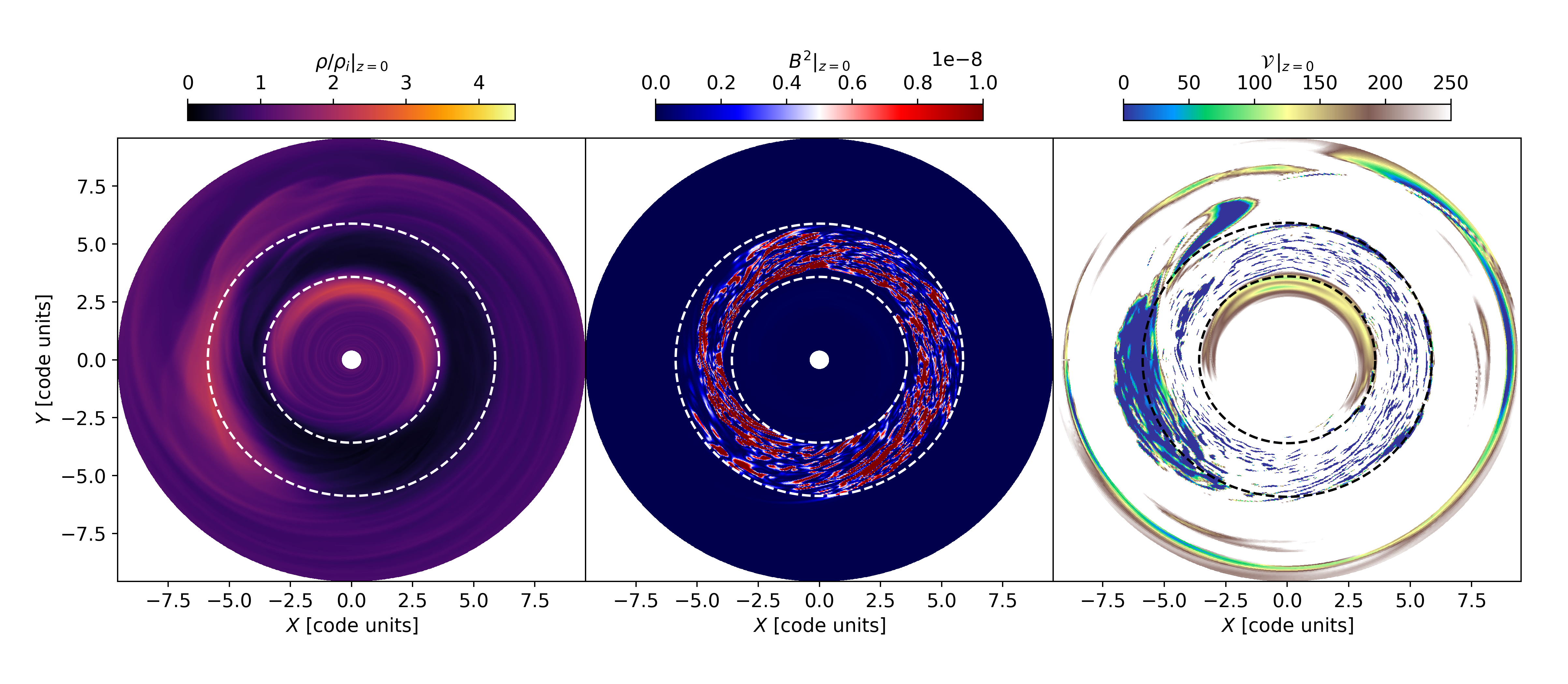}
  \caption{Maps of the normalized gas density $\rho/\rho_i$, squared modulus of the magnetic field $B^2$, and the gas vortensity $\mathcal{V}$ at $z=0$, for the IACT0101b model. Here, the dashed circles delineate the (intermediate) active zone.}
  \label{fig:ActDBV}
\end{figure*}

\subsection{RWI triggered at resistivity transitions invades the dead zone}

Fig. \ref{fig:DBV} shows 2D maps of the normalized gas density $\rho/\rho_0$, the square of the magnetic field $B^2$ and the gas vortensity $\mathcal{V}$, at $t=450$ orbits for models IDEAD0101, IDEAD0801 and IDEAD082c. These models were chosen to explore the effect of both sharp ($H_{b_1}=0.1$) and smooth ($H_{b_1}=0.8$) internal transitions, as well as sharp ($H_{b_2}=0.1$) and very smooth ($H_{b_2}=2$) outer transitions in the diffusivity profile. Note that  the radial extension of the dead zone in model IDEAD0101,
$\Delta r_\mathrm{idz} \equiv r_{2}-r_{1}\simeq 2.7r_0$, is a bit larger than it is in 
model IDEAD0801 ($\Delta r_\mathrm{idz}\simeq 3.1r_0$). In
model IDEAD082c, $\Delta r_{\mathrm{idz}}=5r_{0}$.

In models IDEAD0101 and IDEAD0801 we can see an elongated overdensity in the inner edge of the dead zone, which resembles the shape of a vortex, as well as spiral arms propagating towards the active regions of the disc (see column 1 in Fig. \ref{fig:DBV}). These overdense regions are permeated by a relatively intense magnetic field as can be seen in their magnetic counterparts (see second column of Fig. \ref{fig:DBV}). To verify that these overdensities actually represent the triggering of the RWI, we display the gas vortensity in Fig. \ref{fig:DBV}. In both models the vortensity map indicates that there are two vortices, one  in each edge of the deadzone. It is important to note that in these two models (which have $H_{b_2}=0.1$), the vortex near the outer transition is completely confined to the dead zone. As a consequence, this vortex cannot drag the magnetic field from the outer active zone and no magnetic field counterpart can be seen at the outer edge of the dead zone.

Finally, we consider the model IDEAD082c. Even if the  internal transition is smooth ($H_{b_1}=0.8$), it also exhibits the formation of a vortex similar to that of the model IDEAD0801, which also drags the magnetic field into the dead zone (see third row in Fig. \ref{fig:DBV}). Remarkably, a second elongated vortex at $r\approx 8r_0$ is also formed; we speculate that it arises because of the vortensity gradient in this region (see Section \ref{sec:discussion}).

It is important to note that given that the vortex formation in the IDEAD models takes place within the region where resistivity acts (see Fig. \ref{fig:rprofiles} and the third column in Fig. \ref{fig:DBV}), vortices survive elliptical instability \citep[][]{MB2009,LK2011,MB2012,Lyra_etal2015}, so they persist until the end of our simulations.\footnote{We mention that we have ruled out any change in strength and extension of the vortices in our set-up due to the inclusion of the indirect gas term in the gravitational potential (see Appendix \ref{ap:appendixB}).}

\subsection{Angular momentum transport through an intermediate dead zone}
\label{subsec:amt}

Fig. \ref{fig: aRt} shows the temporal evolution of $\alpha$ Shakura and Sunyaev parameter (see Eq. \ref{eq:alpha}) as a function of radius for the IDEAD0101, IACT0101b, IDEAD0801 and IACT0108 models. Let us focus first on the IDEAD0101 and IDEAD0801 models (that is, with an intermediate dead zone). In the case of the IDEAD0101 model, we find that at $t>30$ orbits, the $\alpha$-parameter of the angular momentum transport increases within the dead zone and its magnitude becomes similar to that of the active zones after $90$ orbits. It should be noted in Fig. \ref{fig: aRt} that the pattern of $\alpha$ within of the dead zone, between $r=3.5$ and $r=6.0r_0$, resembles the shape of a zipper. We argue that this pattern is a consequence of the formation of two vortices (each one at the edges of the dead zone) and of spiral waves emitted by them (see Fig. \ref{fig:dens}). However, we note
an important difference regarding the impact of
the RWI at each edge. The vortex at the inner edge contributes to increase of both components of the stress tensor $\alpha_\mathrm{Rey}$ and $\alpha_\mathrm{Max}$, whereas the vortex formed at the outer edge contributes only to the Reynolds component of the stress tensor $\alpha_\mathrm{Rey}$ (since it does not drag a magnetic field into the dead zone, see for instance Fig. \ref{fig:magnetic}).

On the other hand, for the IDEAD0801 model, we find again a zipper pattern in the $\alpha$-parameter.  This pattern starts to emerge at $t=85$ orbits and is done growing at $t=235$ orbits. During that time interval, $\alpha$ in the dead zone is lower than it is in the active zones. However, at $t=235$ orbits and beyond, $\alpha$ in the dead zone reaches values similar to those in the active zones.

Fig. \ref{fig: aRM} shows $\alpha_\mathrm{Rey}$ and $\alpha_\mathrm{Max}$ for six different models (see Table \ref{tab:initial_conditions}) at $t=450$ orbits. Let us first focus on the left column of that figure, that is, in cases where  the resistivity is given by Eq. (\ref{eq:eta_idead}) and thus the dead zone is in the central part of the disc. On the other hand, the change in the radial extension of the dead zone, $\Delta r_\mathrm{idz}$, does not produce significant changes in the components of the stress tensor. Although apparently a change in $\Delta r_\mathrm{idz}$ leads to an increase of the Reynolds component of the stress tensor $\alpha_\mathrm{Rey}$, that increase is a consequence of the development of the RWI, which may or may not completely cover the dead zone (see above). This can be seen by comparing the models IDEAD082a and IDEAD082c. Note that in these models the value of $r_2$ has been kept fixed. When $\Delta r_\mathrm{idz}=1.5r_0$ (IDEAD082c model), $\alpha_\mathrm{Rey}$ increases (and reaches values similar to those of the active zone $\alpha_\mathrm{Rey}\approx10^{-2}$). However, it should be emphasized that in the region where this increase takes place, there is the formation of a large elongated vortex inside the dead zone (see Fig. \ref{fig:DBV}). Curiously, a similar increase in $\alpha_\mathrm{Rey}\approx10^{-2}$ occurs in the IDEAD082a model only between $r=2.5r_{0}$ and $r=4r_0$, where this also corresponds to the region where the vortex forms. In this model, $\alpha_\mathrm{Rey}$ exhibits a decreasing behavior beyond $r>4r_0$, which is due to the fact that the momentum transport is carried by only the spiral waves emitted by the vortex.

Let's return to model IDEAD0801 to assess the effect of an intermediate value of $\Delta r_\mathrm{idz}$ having a steeper outer transition. In this model, the outer resistivity transition is $H_{b_2}=0.1$. In Fig. \ref{fig: aRM}, it can be seen how the stress tensor component $\alpha_\mathrm{Rey}$ 
increases in the outer part of the dead zone ($r>5.2r_0$) and in the middle of the external active zone ($7.3r_0<r<9.2r_0$). On the other hand, it is important to note that the component of the Maxwell tensor, $\alpha_\mathrm{Max}$, reaches the same magnitude both in the internal and in the external active zones ($\approx10^{-2}$).

Therefore, we find that regardless of the radial extent of the dead zone and the width of the outer transition, $\alpha_\mathrm{Max}$ has a non-zero value in a well-defined region within the dead zone. We identify that this region is where the Rossby vortex instability occurs.

\subsection{Intermediate active zone}

For models with an intermediate active zone (corresponding to $\eta$ as given by Eq. \ref{eq:eta_iact}) we find that as the radial extent of the active zone $\Delta r_\mathrm{iact}=r_2-r_1$ decreases, the value of $\alpha_\mathrm{Rey}$ in the internal dead zone becomes smaller, to the point where it is very low in comparison with the active and the outer dead zones in model IACT0101b (see right-hand column in Fig. \ref{fig: aRM}). However, the magnitude of $\alpha_\mathrm{Rey}$ in the outer dead zone is again governed by the width $H_{b_2}$ of the resistivity transition. The reason is that when the internal transitions are sharper, a very thin and elongated vortex is formed and only weak spiral waves disturb the internal dead zone. On the other hand, for outer transitions, the vortex may extend beyond the intermediate active zone. 

Unlike the IDEAD models, we find that when the vortex penetrates to the intermediate active zone, the magnetic field is removed somehow from the vortex. For instance, Fig. \ref{fig:ActDBV} shows the 2D maps of the normalized gas density $\rho/\rho_i$, the squared modulus of the magnetic field $B^2$ and the gas vortensity $\mathcal{V}$, at $t=450$ orbits for the IACT0101b model. An elongated structure (vortex) can be seen in the gas density close to the internal edge of the active zone, whereas a vortex of greater radial extension is formed at the external edge of the active zone (see left panel in Fig. \ref{fig:ActDBV}). A detailed inspection of the 2D map of the magnetic field (middle panel in Fig. \ref{fig:ActDBV}) reveals that the regions in the active zone where the magnetic field decreases coincide with the location of the vortices (see the vortensity map in Fig. \ref{fig:ActDBV}). 
It is likely that the same expulsion effect of the magnetic field is present in the fiducial simulation of \citet{Lyra_Mac2012}. We think that this effect could be result of an advection process due to the vortex rotation.

It is worth mentioning that, as a consequence of the removal of the magnetic field from the vortex, the component of the Maxwell tensor can become very small (see for instance $\alpha_\mathrm{Max}$ between $r=5.0r_{0}$ and $r=6.5r_0$ in the IACT0108 model presented in Fig. \ref{fig: aRM}). On the other hand, due to the formation of the vortex itself and the formation of spiral waves that propagate in the inner and outer dead zones, we find that $\alpha_\mathrm{Rey}$ in the dead regions can take values from $10^{-4}$ to $10^{-2}$ (see IACT0101a, IACT0101b and IACT0108 models in Fig. \ref{fig: aRM}). Therefore, the value of $\alpha$ in the dead zones is dominated by $\alpha_\mathrm{Rey}$, because $\alpha_{\mathrm{Max}}$ is essentially zero given that the magnetic field is confined to the intermediate active zone.

Fig. \ref{fig: aRt} shows $\alpha$ for the IACT0101b and IACT0108 models. In model IACT0101b,  $\alpha$ exhibits a well-defined pattern in the outer dead zone after $t\approx 80$ orbits, which is caused by the formation of a vortex beyond the active zone. On the contrary, in model IACT0108, the clumpy structure of $\alpha$ in the outer dead zone is almost unnoticeable. In this model, $\alpha$ in the outer dead zone is driven by the spiral waves emitted by a vortex trapped in the active zone.

It should be noted that, since the vortices in the IACT models remain within the active region (where the effect of resistivity can become negligible), they could be subject to elliptical instability \citep[][]{LP2009,LK2011}. However, 
the growth of the Rossby wave instability can prevail over the elliptical instability because the widths of the resistivity transitions used in these models are abrupt \citep{MB2012}. It should also be noted that our simulations have $H/dr\approx 16$ points over the disc 
pressure scale length ($dr$ is the radial size of the cells at $r=r_0$), which is similar to the intermediate resolution used in \citet{Lyra_Mac2012} where they show that vortices can be stable.

\begin{figure}
    \centering
    \includegraphics[scale=0.5]{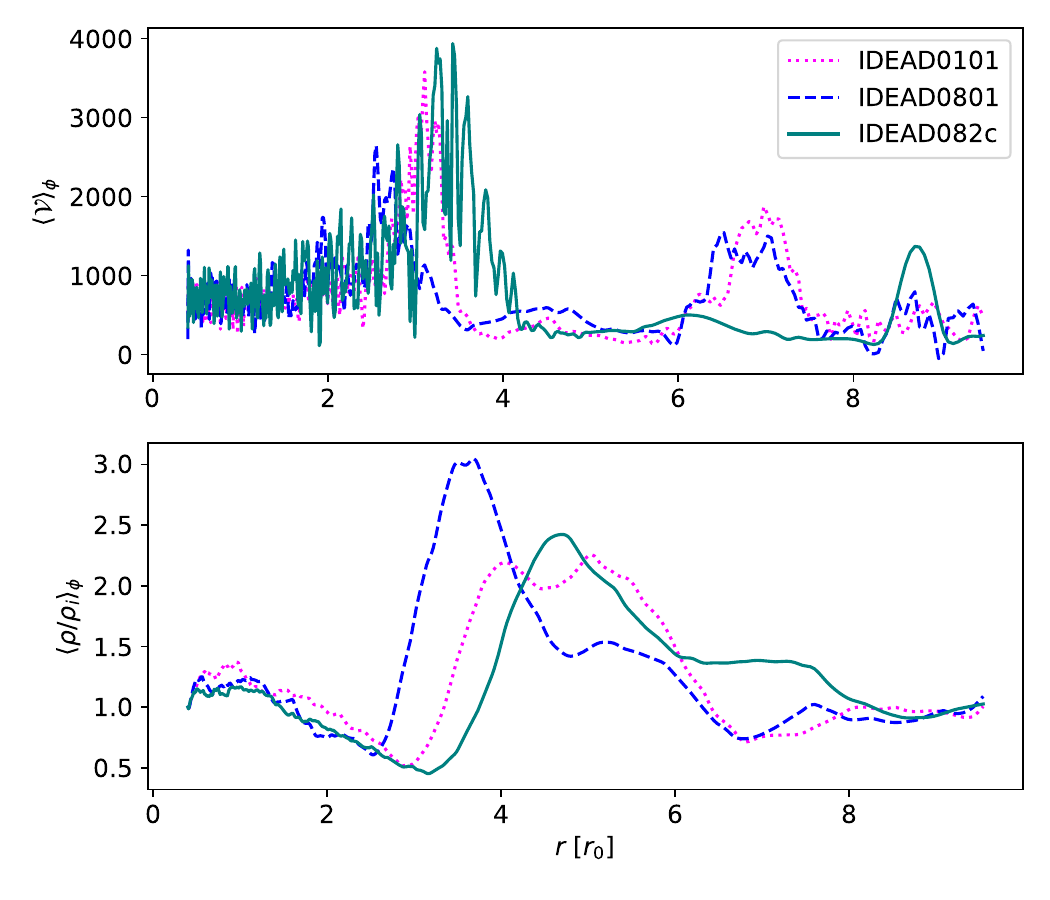}
    \caption{Azimuthally averaged vortensity (\textit{upper panel}) and density (\textit{lower panel}) profiles at the mid-plane for three different models with an intermediate dead zone, at $t=450$ orbits.}
    \label{fig:profiles}
\end{figure}

\section{Discussion}
\label{sec:discussion}

As we showed in the previous section, where a vortex is trapped in an intermediate active zone, the stress tensor component $\alpha_\mathrm{Max}$ disappears because RWI dominates over MRI. In this section we focus on discussing the effects of RWI when there is an intermediate dead zone in the protoplanetary disc.

\subsection{Activation of the dead zone when RWI is present: Magnetic field drag}

The development of RWI in the transition regions between active and dead zones of the disc is considered as one of the main mechanisms that can increase the transport of angular momentum within the dead zone. The formation of vortices in these regions excite spiral waves that disturb the dead zone \citep{Lyra_etal2015,ChCh2023}. Therefore, the resulting turbulence is not a consequence of the development of MRI or of any type of process that involves a magnetic field; these are expected to be highly suppressed due to the low levels of ionization in this region of the disc. However, here we find that in all our models with an intermediate dead zone, regardless of whether the transition of $\eta$ in the inner edge is sharp or smooth, large vortices are formed. Interestingly, these vortices can drag magnetic field lines into the dead zone (see Fig. \ref{fig:magnetic} and middle column in Fig. \ref{fig:DBV}). This drag of the magnetic field lines is mainly driven by the core of the vortex and can considerably increase the component of the Maxwell tensor producing a value of the order of $10^{-3}$. We speculate that this is due to an advection process, which is maintained until the end of our simulations.

\subsection{Comparison with previous studies}

\citet{Ryan2016} studied the role of the RWI in the dead zone by two-dimensional hydrodynamical simulations including viscosity transitions. They found that the development of the RWI can lead to the production of vortices of atypical shapes, depending on the level of viscosity and the radial extension of the dead zone. These vortices and the spiral arms that emerge from them give rise to an increase of the angular momentum transport within the dead zone, with values of $\alpha_\mathrm{Rey}\approx10^{-2}$.  \citet{Ryan2016} concluded that a higher efficiency of angular momentum transport is achieved when the dead zone is narrower because only one vortex is formed or, at most, two coherent antipodal vortices which can come to interact.

Remarkably, in all our models with an intermediate dead zone (IDEAD models), we find the formation of at most two vortices which can interact strongly with each other within the extent of the dead zone (see rows 1 and 2 in Fig. \ref{fig:DBV}) when $H_{b_1}$ and $H_{b_2}$ are small, i.e. when both transitions are sharp. In the case where there is a inner sharp transition and a smooth outer transition, which is the case considered in the IDEAD082c model, we find that only one elongated vortex forms within the dead zone (see Fig. \ref{fig:DBV}). Note that due to the value of the width of the outer transition $H_{b_2}=2.0$ in this model, the MRI is not developed at radii $r>\tilde{r}_2$, which allows us to isolate the effect of the RWI on the $\alpha_\mathrm{Rey}$ stress component in the outer part of the disc.  We obtained in this case a value of $\alpha_\mathrm{Rey}\approx10^{-2}$ in that region of the disc (see Fig. \ref{fig: aRM}), which is in agreement with what was found in \citet{Ryan2016}. It should be noted that this value of $\alpha_\mathrm{Rey}$ is due to the formation of a very elongated vortex as seen in Fig. \ref{fig:DBV} (and confirmed in the vortensity gradient in Fig. \ref{fig:profiles}). When vortices are not formed in the external part of the disc, the value of $\alpha_\mathrm{Rey}$ decreases by an order of magnitude as can be seen in model IDEAD082a in Fig. \ref{fig: aRM}. Under those circumstances, only density waves generated by a vortex formed in the inner transition propagate in the outer parts of the disc.

On the other hand, \citet{Lyra_Mac2012} studied the effect of the RWI on the transport of angular momentum in the dead zone adopting a fixed sharp resistivity transition (that is, $H_{b_1}=10^{-2}$), which is characteristic in the inner zone of the disc. They found that the value of $\alpha_\mathrm{Rey}$ can also reach a value of $\sim 10^{-2}$, which was attributed to the spiral waves emerging from the active zone and from the vortex itself embedded in the side of the dead zone. Although they use a disc model and resistivity function very similar to the one used in our study, their results do not show that the magnetic field is dragged into the dead zone by the vortex.

\subsection{Observed asymmetries in protoplanetary discs}
It has been suggested that the asymmetries observed in transitional discs in millimeter continuum emission \citep[][]{Birnistiel_etal2013,Isella_etal2013,Van_etal2013,Perez_etal2014,Casassus2016}, as is the case of Oph IRS 48, could be vortices triggered by the RWI in
 the transition zones between the active and dead regions of the disc
\citep[e.g.,][]{Lyra_etal2015}. We emphasize that, although our simulated distributions are for the gas and therefore they do not faithfully represent the structures observed in the dust, they do give 
some hints on the dust morphology. For instance, we find patterns of spiral arms in the gas density that propagate beyond the dead zone (see first column in Fig. \ref{fig:DBV}), which resemble the structures observed in several protoplanetary discs \citep{Garufi2013,Grady2013,Benisty2015,Reggiani2018}.

Because in our IACT/IDEAD models vortices are formed in the active/dead transition regions, they could be produced along a considerable radial and azimuthal extension. These vortices
could be very efficient dust traps 
and could facilitate the formation of planetesimals \citep[see][and references therein]{Raetting2021}. However, it must be kept in mind that although the perturbations generated by the vortex and the spiral waves excited in the dead zone, would lead to a greater efficiency in the pebble accretion onto planetesimals (because they prevent the isolation of the planetesimal from the flow of pebbles during the runaway growth), these perturbations can also excite eccentricity and stochastic migration to the planetesimals which can considerably reduce the growth rates of the planetary cores \citep[see][]{Nelson2005}. In other words, the level of turbulence generated in the dead zone by the formation of vortices and spiral waves can change the fate of planetesimals, since their growth rate depends on the turbulent properties of the gas disc \citep{Dzyurkevich_etal2010,OSH2012,OO2013,XB2022}.

Finally, recent interferometric studies of sub-au regions of protoplanetary discs \citep[e.g. HD 163296;][]{Varga_etal2021,SB_etal2021A} suggest that a large-scale time-variable vortex might be present, possibly forming at the transition between the innermost thermally ionized active zone near $\sim1000$ K and an adjacent outer dead zone. \citet{Varga_etal2021} argued that in order to become observable, the vortex must harbour considerably smaller dust grains than the neighbouring disc. In the light of our results, this is indeed possible because the increase of turbulent velocities that we found at the vortex location could in principle decrease the maximum possible size that the dust can grow before fragmenting \citep[$St_\mathrm{frag} \propto \alpha^{-1}$; e.g.][]{Bir_etal2012}. 

\section{Conclusions}
\label{sec:conclusions}

We have performed three-dimensional global magnetohydrodynamical simulations of non-stratified discs including Ohmic resistivity transitions, aimed at checking the angular momentum transport driven by Rossby vortex instability in the otherwise dead zone by varying the widths of the resistivity transitions ($H_{b_1}$, $H_{b_2}$) as well as the radial extent of the dead zone ($\Delta_\mathrm{idz}$).

It is known that the vortex formation at the edges of dead zones is possible even with a shallow resistivity gradient \citep[][]{Lyra_etal2015,ChCh2023} and they can produce an increase in the Reynolds and Maxwell tensors by means of the spiral waves that emerge from them \citep{VT2006,Lyra_Mac2012,Lyra_etal2015}. Here, we found that the activation of the dead zone, that is, the increase in angular momentum transport is mainly a consequence of the vortex formation, and therefore explicitly depends on the resistivity transition widths $H_{b_1}$, $H_{b_2}$ (which in turn depend on the degree of ionization of the gas).

The novelty of our results lies in the fact that the components $\alpha_\mathrm{Max}$ and $\alpha_\mathrm{Rey}$ of the stress tensor increase in the regions of the disc where the vortex is located, reaching values similar to those of the active zones, ranging from $\sim 10^{-3}-10^{-2}$.

Our results are important to understanding angular momentum transport in poorly ionized regions within the disc. In addition, our finding that the vortices emerging within the dead zone carry magnetic field with them, may have interesting implications in the dynamics of the dust grains that are accumulated in the vortices.

\section*{Acknowledgements}
We thank the referee for carefully reading our manuscript and for very useful comments. This work was supported by the Czech Science Foundation (grant 21-11058S).
The work of O.C. was supported by the Charles University Research program (No. UNCE/SCI/023). Computational resources were available thanks to the Ministry of Education, Youth and Sports of the Czech Republic through the e-INFRA CZ (ID:90254).

\section*{Data Availability}

 The FARGO3D code is available from \href{https://fargo3d.bitbucket.io/intro.html}{https://fargo3d.bitbucket.io/intro.html}. The input files for generating our 3D magneto-hydrodynamical simulations will be shared on reasonable request to the corresponding author.



\bibliographystyle{mnras}
\bibliography{manuscript} 




\appendix
\section{Initial magnetic field configuration}
\label{ap:appendixA}

\begin{figure*}
    \includegraphics[scale=0.5]{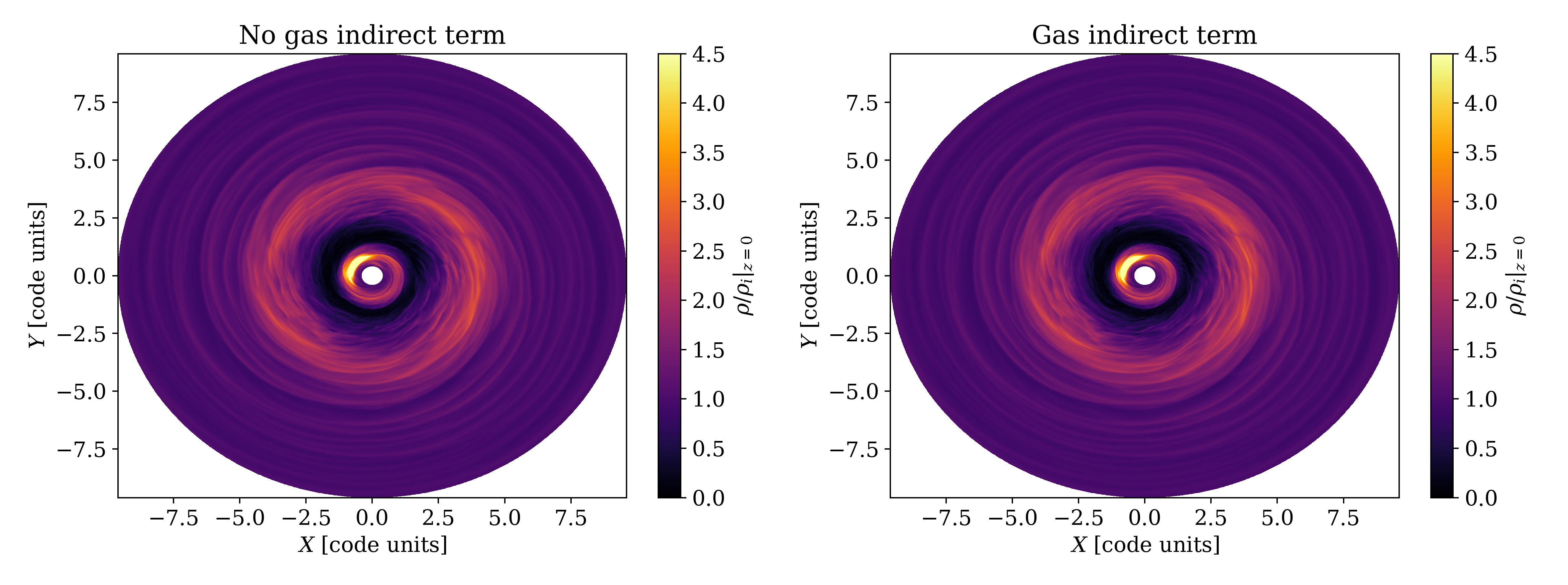}
  \caption{Maps of the normalized gas density, $\rho/\rho_i$, for two models, with and without gas indirect term, after $t=250$ orbits (see Appendix \ref{ap:appendixB} for details).}
  \label{fig:it}
\end{figure*}

In all our numerical models the magnetic field is set as a net vertical field \citep[see][and references therein]{ChCh2023}, whose initial configuration in non-stratified discs implicitly depends on the vertical box length 
$L_z$. Since the excitation of MRI takes place if the most unstable wavelength $\lambda_{\mathrm{MRI}}$ is contained within the vertical domain of the disc \citep[][]{BH1991}, 
we demand that two MRI wavelengths are resolved in the vertical height of our numerical simulations. That is,

\begin{equation}
2\lambda_\mathrm{MRI}=\frac{4\pi v_A}{\Omega_\mathrm{Kep}(r)}=L_z,
    \label{eq:lamb_MRI}
\end{equation}
where $\Omega_\mathrm{Kep}(r)=\Omega_0(r/r_0)^{-3/2}$. From this condition we can infer the initial magnitude of the vertical
magnetic field versus radius.

Equation (\ref{eq:lamb_MRI}) implies that
\begin{equation}
v_A=\frac{\Omega_0 L_z}{4\pi}\left(\frac{r}{r_0}\right)^{-\frac{3}{2}}.
    \label{eq:Va}
\end{equation}
Recalling that the Alfvén speed is defined as $v_{A}\equiv B/\sqrt{\mu_{0} \rho}$ and equating Eqs (\ref{eq:lamb_MRI}) and (\ref{eq:Va}), we obtain that 
\begin{equation}
B=\frac{\Omega_0 (\mu_{0} \rho_0)^{1/2} L_z}{4\pi}\left(\frac{r}{r_0}\right)^{-\frac{3}{2}}\left(\frac{r}{r_0}\right)^{-\frac{q_\rho}{2}}.
    \label{eq:B_r}
\end{equation}
Here we have also used that the initial radial profile of $\rho$ is given by Eq. (\ref{eq:rho}). Finally, grouping all the constants within $B_0$, we obtain that the initial profile of the magnetic field is given by
\begin{equation}
B=B_0\left(\frac{r}{r_0}\right)^{-(\frac{3+q_\rho}{2})}.
    \label{eq:Bfield_prof}
\end{equation}

\section{On the effect of the indirect gas term}
\label{ap:appendixB}

In numerical simulations where the reference frame is centered on the star instead of the center of mass of the system, it is common
to include the "\textit{indirect term}" in the moment equation (see Eq. \ref{eq:momentum}) to account for the non-inertial term
due to the gravitational potential generated by the mass distribution of the disc \citep[e.g.,][]{Regaly_etal2013}. \citet{ZB2016} found that for massive discs (those with $Q<\pi/2h$, where $Q$ is the Toomre parameter and
$h$ the disc's aspect ratio), vortices are much weaker when both the indirect term and disc self-gravity are included 
than in simulations with only the indirect term \citep[see also][]{Regaly_etal2017}. \citet{Crida_etal2022} point out that if disc self-gravity is not taken into account, then the 
indirect should be also ignored to avoid spurious destabilization of the disc. It remains unclear how the level of destabilization depends on
whether the model is 2D or 3D.

In order to assess how much the indirect term affects the results,
we carry out two simulations, with and without taking into account the indirect term. Recall that our models are unstratified 3D discs where the gravitational 
potential does not depend on $z$. In these two experiments we use a very smooth resistivity transition in the computational domain, which is equivalent to an extended intermediate dead zone (that is, IDEAD type):
$\tilde{r}_1=5$, $\tilde{r}_2=10$, $H_{b_1}=2$ and $H_{b_2}=0.1$. 
Figure \ref{fig:it} shows the gas density at the disc mid-plane in both simulations. As we see, both simulations give similar results. 
In fact, the density changes by less than $1$ percent between both runs. The vortex formed at $r\approx1$ is essentially not affected. Therefore, we conclude that at least in 
our non-stratified 3D disc models, the gas indirect term is unimportant.



\bsp	
\label{lastpage}
\end{document}